\documentclass[10pt,conference]{IEEEtran}

\usepackage{amsmath,amssymb,amsfonts}
\usepackage{xcolor}
\usepackage{tabularx}
\usepackage{hyperref}
\usepackage{xspace}
\usepackage{mdframed}
\usepackage{multirow}
\usepackage{makecell}
\usepackage{booktabs}
\usepackage{array}
\usepackage{siunitx}
\usepackage{cite}
\usepackage{pgfplots}
\usepgfplotslibrary{statistics}
\usepackage{threeparttable}
\usepackage[disable]{todonotes}
\usepackage{listings}
\usepackage[inline]{enumitem}

\definecolor{callout}{RGB}{0,120,120}      %

\makeatletter

\newcommand{\changed}[1]{}%

\newcommand{\etal}{et al.\xspace}
\newcommand{\ie}{i.e.,\xspace}
\newcommand{\eg}{e.g.,\xspace}
\newcommand{\fig}[1]{Figure~\ref{#1}}
\newcommand{\tab}[1]{Table~\ref{#1}}
\newcommand{\sect}[1]{Section~\ref{#1}}

\newcommand{\cicd}{CI/CD\xspace}
\newcommand{\github}{GitHub\xspace}

\newcommand{\gha}{GHA\xspace}
\newcommand{\ghaw}{GHA workflow files\xspace}

\newcommand{\wfs}{workflows\xspace}

\newcommand{\fr}{\emph{Failure Rate}\xspace}
\newcommand{\commits}{\emph{\#Commits}\xspace}
\newcommand{\aval}{\emph{Availability}\xspace}
\newcommand{\mttr}{\emph{MTTR}\xspace}

\definecolor{listingbg}{HTML}{FBFBFD}
\definecolor{listingrule}{HTML}{D1D5DB}
\definecolor{yamlkeyword}{RGB}{0,95,150}
\definecolor{yamlstring}{RGB}{206,100,23}
\definecolor{yamlcomment}{RGB}{120,120,120}
\definecolor{callout}{RGB}{0,120,120}

\definecolor{gray}{cmyk}{0,0,0,0.50}

\lstdefinelanguage{yaml}{
  sensitive=false,
  morecomment=[l]{\#},
  morestring=[b]",
  alsoletter={-:\/\.\$},
 morekeywords={},
}

\lstdefinestyle{yamlstyle}{
  language=yaml,
  basicstyle=\ttfamily\footnotesize,
  breaklines=true,
  columns=fullflexible,
  keepspaces=true,
  showstringspaces=false,
  backgroundcolor=\color{listingbg},
  rulecolor=\color{listingrule},
  commentstyle=\color{yamlcomment}\itshape,
  stringstyle=\color{yamlstring},
  keywordstyle=\color{yamlkeyword}\bfseries,
  frame=single,
  framerule=0.4pt,
  numbers=left,
  numberstyle=\ttfamily\tiny,
  numbersep=5pt,
  xleftmargin=1.5em,
  framexleftmargin=1.5em,
  xrightmargin=2pt,
  aboveskip=6pt,
  belowskip=6pt,
  literate=
    {name:}{{\color{yamlkeyword}name:}}1
    {on:}{{\color{yamlkeyword}on:}}1
    {push:}{{\color{yamlkeyword}push:}}1
    {permissions:}{{\color{yamlkeyword}permissions:}}1
    {contents:}{{\color{yamlkeyword}contents:}}1
    {branches:}{{\color{yamlkeyword}branches:}}1
    {jobs:}{{\color{yamlkeyword}jobs:}}1
    {build:}{{\color{yamlkeyword}build:}}1
    {test:}{{\color{yamlkeyword}test:}}1
    {strategy:}{{\color{yamlkeyword}strategy:}}1
    {matrix:}{{\color{yamlkeyword}matrix:}}1
    {os:}{{\color{yamlkeyword}os:}}1
    {runs-on:}{{\color{yamlkeyword}runs-on:}}2
    {steps:}{{\color{yamlkeyword}steps:}}1
    {uses:}{{\color{yamlkeyword}uses:}}1
    {run:}{{\color{yamlkeyword}run:}}1
    {needs:}{{\color{yamlkeyword}needs:}}1,
}

\newcommand{\anonymise}[1]{<anonymised>}

\newcommand{\summary}[1]{\begin{mdframed}[backgroundcolor=black!5, topline=false, rightline=false,  leftline=true, bottomline=false,  linewidth=2pt, linecolor=black!70, nobreak=false]\noindent{#1}\end{mdframed}}

\global\mdfdefinestyle{exampledefault}{linewidth=1pt, linecolor=lightgray!80, backgroundcolor=lightgray!20, innerleftmargin=4pt, innerrightmargin=4pt, innertopmargin=3pt, innerbottommargin=3pt, nobreak=true,roundcorner=5pt}
\newenvironment{custombox}{\smallskip\begin{mdframed}[style=exampledefault]}{\end{mdframed}\smallskip}

\def\BibTeX{{\rm B\kern-.05em{\sc i\kern-.025em b}\kern-.08em
    T\kern-.1667em\lower.7ex\hbox{E}\kern-.125emX}}

\begin{document}

\title{On the GitHub Actions Language:\\
Usage, Evolution, and Workflow Reliability}

 \author{
   \IEEEauthorblockN{Aref Talebzadeh Bardsiri}
   \IEEEauthorblockA{\textit{Software Engineering Lab}\\
     \textit{University of Mons}\\
     Mons, Belgium\\
     aref.talebzadehbardsiri@umons.ac.be}
 \and
   \IEEEauthorblockN{Alexandre Decan}
   \IEEEauthorblockA{F.R.S.-FNRS Research Associate\\
     \textit{Software Engineering Lab}\\
     \textit{University of Mons}\\
     Mons, Belgium\\
     alexandre.decan@umons.ac.be}
 \and
   \IEEEauthorblockN{Tom Mens}
   \IEEEauthorblockA{\textit{Software Engineering Lab}\\
     \textit{University of Mons}\\
     Mons, Belgium\\
     tom.mens@umons.ac.be}
 }

\maketitle

\begin{abstract}
Developers often struggle with maintaining GitHub Actions workflow configurations in GitHub-hosted repositories, with recent studies showing %
frequent execution failures.
This paper empirically explores
how the adoption and evolution of GitHub Actions language constructs impacts workflow reliability and maintainability.
To do so, we quantitatively analyse  260K workflows from 49K GitHub repositories to understand how they are used in practice and how their usage has evolved from July 2019 to August 2025. %
We identify 197 language constructs available in the GitHub Actions language
and map them to 14 features reflecting workflow capabilities.
We observe that only a small set of constructs and features are used very frequently, and
that larger and more complex workflows are associated with higher failure rates and more maintenance effort. 
We identify specific features that are more likely to be linked with reliability and maintainability risks.
These insights can help  practitioners and researchers improve their understanding and usage of the GitHub Actions language, which can help in improving and sustaining workflow automation practices.%
\end{abstract}


\IEEEpeerreviewmaketitle

\section{Introduction}\label{sec:introduction}

With the increasing demand for efficient and high-quality software systems,
\cicd practices have become mainstream in software projects to streamline their development and deployment pipelines.
\cicd services automate repetitive tasks such as building code, running tests, performing quality and security checks, and deploying applications.
They are an integral part of collaborative software development because they enhance productivity, improve efficiency, and reduce the likelihood of human errors \cite{Zampetti2021}.

In the past, different \cicd services (\eg Travis, CircleCI and Jenkins) were frequently used in \github repositories. Since the public release in 2019 of \github's integrated \cicd solution called GitHub Actions (hereafter shortened to \gha), it has become the most popular \cicd tool on GitHub \cite{Golzadeh2022SANER}.
\gha allows repository maintainers to automate workflows through YAML-based configuration files.
For writing these workflows, \gha provides a rich set of language constructs (\ie\ keys, structures, values, etc.).\footnote{\url{https://docs.github.com/en/actions/reference/workflows-and-actions/workflow-syntax}}
It can be considered as a domain-specific language~\cite{mernik2005DSL} that allows workflow maintainers to define workflows, jobs, steps, and more.

Its seamless integration with GitHub, its large marketplace of Actions, and its generous free plan for running workflows for public repositories, have made \gha a compelling choice for many developers \cite{Golzadeh2022SANER, decan2022gha}.
However, developers have highlighted that using \gha workflows comes with multiple challenges~\cite{Saroar2023DevelopersPO, Onsori2026}, such as the difficulty in writing, testing, and debugging workflow files.
Ghaleb \etal \cite{ghaleb2025complexity} reported that \gha workflows are among the most complex \cicd automation services due to their high maintenance effort, while Zheng \etal \cite{zheng2025ghafail} observed that \ghaw frequently fail during execution, highlighting challenges related to reliability and efficiency.
These observations suggest that \gha syntax and semantics may be poorly understood or mastered by workflow maintainers, or that tool support is below par.

Despite the widespread adoption of \gha, %
few %
 studies have empirically analyzed the language constructs used in \ghaw, their usage patterns, and their evolution over time.
While the aforementioned challenges suggest that complex \cicd configurations have a negative impact on software quality, there is a lack of quantitative
evidence at scale linking the specific structural growth of \ghaw to their reliability and maintainability. Addressing this gap is crucial to help practitioners manage configuration complexity and establish concrete best practices for \cicd automation.

This paper therefore presents a large-scale quantitative study of the \gha language, its usage, and its direct impact on workflow maintainability and reliability. We leverage a dataset of 260K workflows from 49K GitHub repositories, tracking their evolution over a six-year time span from July 2019 to August 2025.
We conduct statistical analyses on a subset of these workflows to evaluate how their structural size and feature usage correlate with reliability and maintainability issues. We aim to answer the following research questions:
\begin{description}[leftmargin=!]

    \item[\textbf{\boldmath $RQ_0$}]\textit{What are the constructs and features of the \gha language?}
    A first step towards understanding the usage of \gha is to identify the constructs provided by its language.
    We show that \gha provides a rich language of 197 distinct constructs, which we group into 14 high-level features.

    \item[\textbf{\boldmath $RQ_1$}] \textit{Which constructs are used in practice?}
    Understanding the usage frequency of the language constructs can reveal which of them are central to workflow configurations and which ones tend to be more specialized.
    We find that only a subset of them are frequently used, that most workflows tend to use a wide variety of constructs, and that several constructs are being used repeatedly within a same workflow.

    \item[\textbf{\boldmath $RQ_2$}] \textit{Which features are used in practice?}
    Similarly, we identify which features are commonly used and which ones are rarely observed in practice.
    In addition, we analyze to what extent workflows make use of each feature in terms of its available constructs, and which constructs are most frequently employed in workflow configurations.

    \item[\textbf{\boldmath $RQ_3$}] \textit{How does the GHA language usage evolve over time?}
    We analyze how the usage of \gha constructs and features has changed during the six-year observation period.
    We find that the number of paths in workflows tends to increase over time by a factor of two to three.
    In contrast, the number of constructs used in workflows only increases slightly, and the number of features used in workflows tends to remain stable.
    This suggests that increases in workflow size are not due to the usage of more features, but rather to the reuse of the same set of constructs in more places.

    \item[\textbf{\boldmath $RQ_4$}] \textit{To what extent does workflow size impact reliability and maintainability?}
    We analyze the relationship between workflow size and workflow reliability and maintainability. We show that larger workflows are significantly more prone to execution failures, require more maintenance effort, have lower overall availability, and take longer to repair after a failure.
    In addition, we show that the use of specific features is more strongly associated with reliability and maintainability issues than other features. %
    \end{description}

The remainder of this article is structured as follows. \sect{sec:background} introduces the core concepts and terminology used in the paper.
\sect{sec:methodology} explains the data extraction. Sections~\ref{sec:rq0} to \ref{sec:rq3} address the research questions.
\sect{sec:related_work} presents the related work.
\sect{sec:threats} discusses the threats to validity, and \sect{sec:conclusion} concludes the paper and discusses the findings.

\section{Background}\label{sec:background}

Like any CI/CD service, %
\gha enables \github repository maintainers to configure workflows to automate the building, testing, analysis and deployment of their software projects.
Developers can also use \gha to automate many other activities, such as managing issues and pull requests, sending notifications, and more.\footnote{\url{https://docs.github.com/en/actions/get-started/understand-github-actions}} Its tight integration with GitHub makes \gha a popular choice among developers~\cite{decan2022gha}.

\gha workflow configurations use the YAML file format and are stored in the \emph{.github/workflows} directory of a \github repository.
Figure~\ref{fig:gha-example} provides an example of a workflow configuration to automate the building and testing of a Node.js project.
Each workflow has a \textsf{name}, which can either be defined in the file (line 1) or inferred from the filename, if not explicitly defined.
Workflows can be triggered by a wide range of \emph{events} (e.g., push, pull request, schedule) specified by the \textsf{on} key (line 2).
Lines 3-4 declare that the workflow is triggered by a \textsf{push} event on the main branch.
The workflow defines two \emph{jobs} labeled \textsf{build} (lines 8-15) and \textsf{test} (lines 16-20). %
Jobs can execute in parallel on one or more \emph{runners} (\textsf{runs-on}, lines 12 and 18) in \github-hosted or self-hosted virtual environments.
Each job declares one or more \textsf{steps} (lines 13-15 and lines 19-20).
A \textsf{run} step (lines 15 and 20) allows maintainers to execute a sequence of shell commands.
A \textsf{uses} step (line 14) executes a reusable component, called \emph{Action}, sourced from public \github repositories and often found on the GitHub Marketplace.\footnote{\url{https://github.com/marketplace?type=actions}} Steps can even reuse entire workflows.\footnote{\url{https://docs.github.com/en/actions/sharing-automations/reusing-workflows}} %
One can define different configurations for a job through the \emph{matrix strategy} (\textsf{strategy} and \textsf{matrix} on lines 9-11), allowing the same job to be run the same job in different environments, operating systems or language versions (\eg the user-defined \textsf{os} variable defined on line 11 and used on line 12).
Maintainers can also specify the \emph{permissions} granted to a workflow (line 5), controlling its access to the repository content (line 6), to the issue tracker, to pull requests, etc. %

\begin{figure}[h]
  \footnotesize
\begin{lstlisting}[style=yamlstyle]
name: CI
on:
  push:
    branches: [main]
permissions:
  contents: read
jobs:
  build:
    strategy:
      matrix:
        os: [ubuntu-latest, windows-latest]
    runs-on: ${{ matrix.os }}
    steps:
      - uses: actions/checkout@v3
      - run: npm ci
  test:
    needs: build
    runs-on: ubuntu-latest
    steps:
      - run: npm test
\end{lstlisting}
\caption{Example of a \gha workflow configuration.}
\label{fig:gha-example}
\end{figure}

The YAML format of workflow configurations imposes a hierarchical structure of key-value pairs, lists, and nested elements.
A \emph{path} in a workflow refers to a specific element within this structure, starting from a top-level key. We separate the different levels of the hierarchy by dots.
For example, the path \textit{\lstinline|jobs.build.steps[0].uses|} refers to the Action used by the first step (at index 0) of the \emph{build} job (line 14 in Figure~\ref{fig:gha-example}).

The \gha syntax allows for user-defined keys in various places.
Figure~\ref{fig:gha-example} shows two examples of user-defined keys for jobs (\textsf{build} on line 8 and \textsf{test} on line 16) and for matrix variables (\textsf{os} on line 11). User-defined keys can also be used for environment variables, Action parameters, etc.\footnote{We refer to the online documentation of \gha for more details.}

To facilitate analysing and comparing paths across different workflows using different user-defined keys, we introduce the concept of \emph{constructs}.
A \emph{construct} represents an abstract workflow path where user-defined keys and list indices are replaced by generic placeholders.
For example, the concrete path \textit{\lstinline|jobs.build.steps[0].uses|} is abstracted to the construct \textit{\lstinline|jobs.<id>.steps[*].uses|}, where \textit{\lstinline|<id>|} is a placeholder for the job identifier and \textit{\lstinline|[*]|} abstracts the actual index of the step.

\section{Data Extraction}\label{sec:methodology}
In order to study \gha language usage evolution, we need a large collection of workflow histories belonging to a diverse set of software development repositories on GitHub, excluding experimental or personal repositories. Cardoen \etal~\cite{Cardoen2024} provide such a dataset, containing the history and content of 267K workflows from 49K software development repositories, covering the period from July 2019 to August 2025, and containing over three million workflow snapshots. The repositories have been selected based on their popularity (at least 100 stars), activity (at least 300 commits) and recency (at least one commit after August 25th, 2024).
We relied on version 2025-10-09 of this dataset to conduct our study.

Since our focus is on the actual usage of the \gha language, we excluded
152,380 \emph{invalid} workflow snapshots (corresponding to invalid YAML files, or containing paths that are not supported by the \gha language).
To do so, we used the validity flag provided in the dataset, %
indicating whether a workflow file conforms to the JSON Schema for \gha workflows.\footnote{\url{https://www.schemastore.org/github-workflow.json}}
A manual inspection revealed that some \emph{a priori} valid workflow snapshots contained \textit{constructs} not supported by the \gha language.
We manually examined all unique constructs extracted from the workflows, and compared them with the official \gha documentation~\cite{githubActionsDocs}. Constructs not mentioned in the documentation were considered invalid, and the 20,750 snapshots containing them were excluded.

After this postprocessing phase, we obtained a final dataset of 2,847,199 workflow snapshots corresponding to 259,661 workflows from 48,952 GitHub repositories, containing 186,706,963 paths.

\section{{$RQ_0$}: What are the constructs and features of the \gha language?}\label{sec:rq0}

Before being able to analyse how \gha is used in practice, we provide a high-level view of the constructs and features that the language provides.
As a first step, we used the official \github documentation~\cite{githubActionsDocs} to identify the constructs provided by \gha.
Given that this documentation is scattered across multiple webpages, manually collecting all language constructs is challenging~\cite{zheng2025ghafail},
so we might have missed some constructs in doing so.

Therefore, as a proxy, we parsed all \gha workflows in our dataset to extract the language constructs.
Since our dataset spans multiple years and the \gha language has evolved over time, some detected constructs that existed previously might have been removed from the documentation
(\eg \textsf{permissions.repository-projects}) and new constructs may have been added (\eg \textsf{permissions.attestations}).
Therefore, we only considered the \changed{latest snapshots of workflows still existing in August 2025 (\ie excluding deleted workflows)}.

This involved 6.5M+ paths from 171,194 workflow snapshots, enabling us to identify 197 valid unique language constructs, of which 119 are workflow-level, 65 are job-level, and 13 are step-level constructs.
This set of snapshots is also used to answer $RQ_1$ and $RQ_2$.

\begin{figure}[!t]
  \centering
  \includegraphics[width=1.05\columnwidth,keepaspectratio,trim={1.6cm 0.5cm 1cm 2cm}]{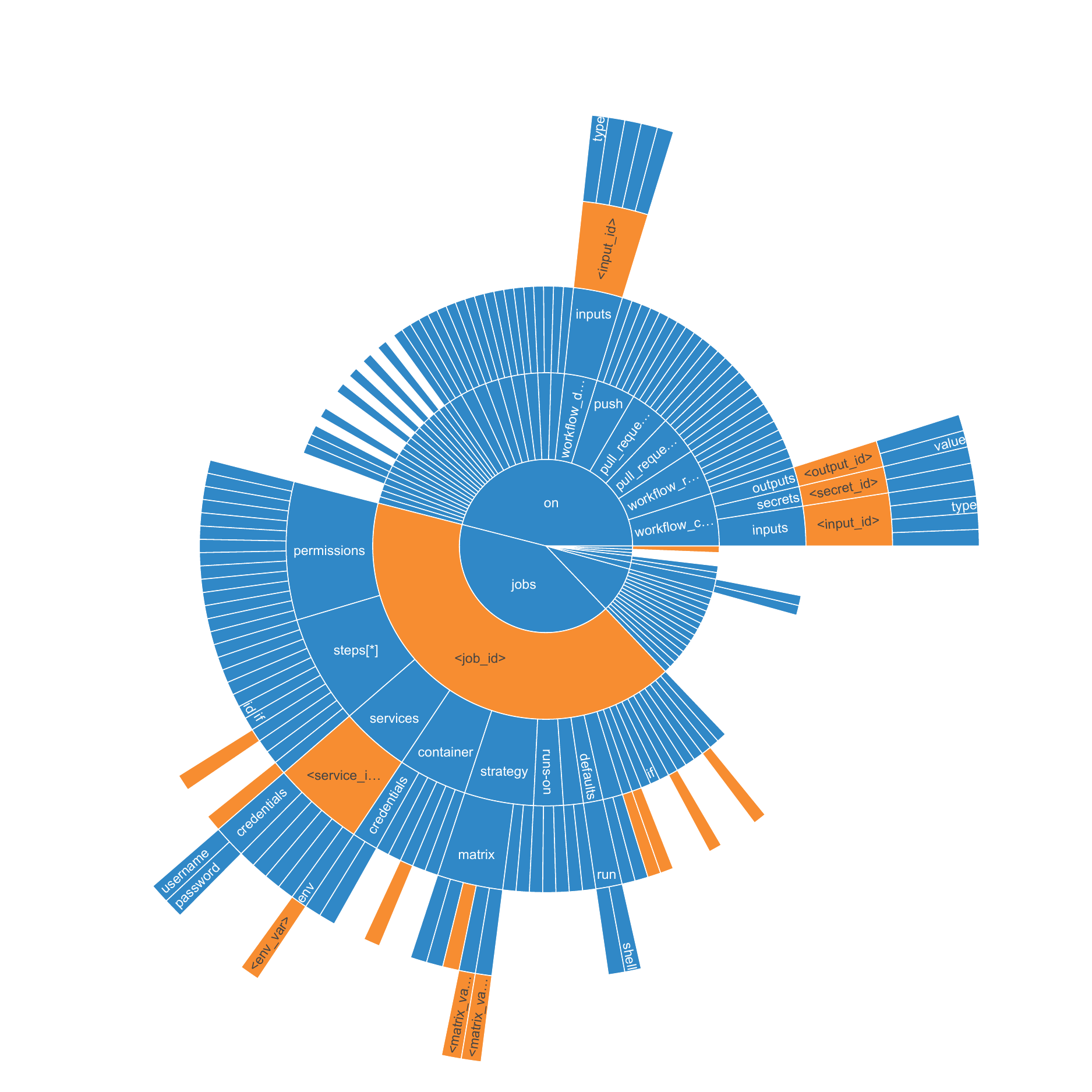}
  \caption{Sunburst diagram of \gha language constructs. Blue segments are predefined keys in workflows, orange segments are user-defined keys.}
  \label{fig:sunburst}
  \vspace{-0.5cm}

\end{figure}

The sunburst diagram in Figure~\ref{fig:sunburst} shows the hierarchical structure of the identified constructs.
For readability purposes some construct names were hidden in this figure, but they are available in an interactive version of the diagram in the replication package. %
One can easily observe that predefined workflow keys (blue segments) are much more abundant than user-defined keys (orange segments). Each segment's size is proportional to the number of sub-keys it contains.
The inner ring of the diagram represents top-level keys such as \textsf{jobs}, triggers (\textsf{on} key), workflow-level \textsf{permissions}, and \textsf{env}ironment variables.
The subsequent rings represent constructs at deeper nesting levels.
The figure reveals that the \gha language contains lots of constructs at different levels in the hierarchy, especially for the workflow triggers (\textsf{on}) and the individual jobs. This may lead to cognitive load for maintainers to learn to use such constructs effectively.
In addition, the many orange segments in Figure~\ref{fig:sunburst} reveal that \gha enables workflow maintainers to extend the language with user-defined constructs, such as custom \textsf{environment variables}, parameters, or job names. While this adds flexibility to the language, these variation points add an extra layer of complexity for maintainers to learn and use the language effectively.

\bigskip
\summary{The rich hierarchically-structured language offered by \gha is a double-edged sword: while providing flexibility and expressiveness to define complex workflows, it may increase the cognitive load for maintainers to learn and use the language effectively.}

Construct-level analysis alone is too granular for understanding the broader capabilities of \gha. Therefore, we grouped and mapped all semantically-related constructs into higher-level \textit{features} reflecting workflow capabilities.
For example, the constructs \textit{\lstinline|jobs.<id>.steps[*].run|} and \textit{\lstinline|defaults.run.shell|} both relate to running commands, so they were mapped to a \textsf{commands} feature encompassing the seven different constructs.
Similarly, the construct \textit{\lstinline|jobs.<id>.runs-on|} relates to the execution context of a job, so it was mapped to the \emph{context} feature along with four other constructs.

\begin{table}[!t]
\centering
\begin{threeparttable}
\caption{List of \gha language features and number of constructs (197 in total) mapped to each feature.}
\label{tab:features}
\footnotesize
\begin{tabular}{l r p{0.45\columnwidth}}

\textbf{Feature} & \textbf{\# Constructs} & \textbf{Example Construct} \\
\midrule

triggers& 85 &

\textit{\lstinline|on.push.branches[*]|} \\

permissions & 30 &
{\textit{\lstinline|permissions.contents|}}\\

workflow reuse & 14 &
{\textit{\lstinline|jobs.<id>.uses|}}\\

job orchestration & 12 &
{\textit{\lstinline|jobs.<id>.if|}}\\

containers & 9 &
{\textit{\lstinline|jobs.<id>.container.image|}}\\

matrix strategy & 8 &
{\textit{\lstinline|jobs.<id>.strategy.matrix.<var>|}}\\

commands & 7 &
{\textit{\lstinline|jobs.<id>.steps[*].run|}}\\

services & 7 &
{\textit{\lstinline|jobs.<id>.services.<s_id>.image|}}\\

environment vars & 6 &
{\textit{\lstinline|jobs.<id>.steps[*].env.<var>|}}\\

naming & 5 &
{\textit{\lstinline|name|}}\\

context & 5 &
{\textit{\lstinline|jobs.<id>.runs-on|}}\\

Action reuse & 3 &
{\textit{\lstinline|jobs.<id>.steps[*].uses|}}\\

step orchestration & 3 &
{\textit{\lstinline|jobs.<id>.steps[*].if|}}\\

deployment & 3 &
{\textit{\lstinline|jobs.<id>.environment|}}\\
\bottomrule

\end{tabular}
\begin{tablenotes}[para,flushleft]
\end{tablenotes}
\end{threeparttable}
\vspace{-0.5cm}

\end{table}

Since \gha does not provide an official taxonomy of such features, we defined our own taxonomy based on a manual analysis of the constructs and the documentation. This process was iteratively conducted by the authors until consensus was reached.
The final mapping consists of 197 constructs grouped into 14 features.
Table~\ref{tab:features} shows the list of features and the number of constructs mapped to each feature, along with an example construct for each feature.
The full mapping is available in the replication package.
Table~\ref{tab:features} reveals that the number of constructs per feature ranges from 3 to 85.
With 85 constructs, feature \emph{triggers} is the most populated because \gha provides a wide range of events to trigger workflows, each with its own set of constructs to configure the trigger's behavior.
The \emph{permissions} feature contains 30 constructs, reflecting the importance of access control in workflows.
The ability of \gha workflows to reuse pre-existing workflows or Actions is reflected by the features \emph{workflow reuse} and \emph{Action reuse}, containing 14 and 3 constructs, respectively.
Along with \emph{Action reuse}, the features \emph{step orchestration}, and \emph{deployment} also contain only three constructs.
\newline

\summary{GitHub Actions offers many features that vary in number of constructs. The cognitive burden on workflow maintainers may depend on the kind of features they need to use. Features that contain more constructs or that are less common may be more difficult to master.}
\section{{$RQ_1$}: Which constructs are used in practice?}\label{sec:rq1a}

Using the \gha constructs identified in RQ$_0$, we examine how their usage varies across the same 171,194 workflow snapshots.

\noindent\textbf{\emph{\#Paths and \#Constructs.}} A distribution analysis of the number of paths and constructs per workflow reveals that workflows typically contain many paths but fewer constructs. For instance, \emph{\#Paths} ranges between 2 and 7,712, with a median of 24, while \emph{\#Constructs} ranges between 2 and 48, with a median of 11. Half of the workflows have between 15 (Q1) and 42 (Q3) paths and between 9 (Q1) and 14 (Q3) constructs. This already indicates that some constructs tend to be repeated multiple times within workflows.

\noindent\textbf{\emph{Inequality in construct usage.}} To understand how construct usage is distributed across workflows, we computed the Gini coefficient, which is a measure of inequality in a distribution~\cite{gini1912Variabilit}.
A Gini coefficient of 0 indicates perfect equality (all constructs are used equally), while a coefficient close to 1 indicates high inequality (a few constructs dominate the usage). We obtained a Gini coefficient of 0.84, indicating strong inequality in construct adoption across workflows.
In fact, the 10 most frequently used constructs account for 58.1\% of all construct occurrences across our dataset, and the top 50 account for 92.9\%. The remaining 147 constructs (74.6\%) account for only 7.1\%.

Next, we examine how many constructs and paths workflows typically use.
A Spearman rank correlation of $\rho = 0.78$ ($p<0.001$) reveals that
the \emph{\#Paths} generally increases with \emph{\#Constructs}.
Still, many workflows have many paths but few constructs, indicating repeated use of the same constructs.
As can be observed from the segment sizes in \fig{fig:sunburst}, repetition can occur through user-defined keys, multiple jobs, and multiple steps.
Overall, workflows vary in diversity (more distinct constructs) and repetition (repeated use of the same constructs).

\noindent\textbf{\emph{Most popular constructs.}} To understand which constructs are most popular in workflows, we analysed the frequency of construct usage across all workflows.
Table~\ref{top_20} reports the top 15 %
constructs based on workflow usage frequency (\%wf) alongside the total number of observed occurrences (\# column). The top construct is the non-mandatory workflow \textit{\lstinline{name}}, suggesting maintainers are likely to provide a name for their workflow.
The second most frequent construct is  \textit{\lstinline{runs-on}} to specify the runner to execute the job.
The overwhelming majority of workflows \textbf{not} using this construct instead relied on a reusable workflow (through \textit{\lstinline{jobs.<id>.uses}}, not part of the top 15%
) that was used to specify the runner.
Seven of the fifteen %
top constructs occur inside individual steps (\textit{\lstinline{jobs.<id>.steps[*]...}}).
For example, the third and fourth top constructs have to do with reusing existing Action components:  \textit{\lstinline{uses}} specifies the Action to be used, and \textit{\lstinline{with.<param>}} provides input parameters to the Action.
This reveals that reusing Actions in steps is common practice (94.2\% of workflows), much more so than providing custom scripts using the \textit{\lstinline{run}} key (ranked 6th, 77.2\% of workflows).
Constructs related to defining workflow triggers (\textit{\lstinline{on}} key) are also quite common. Out of the 34 trigger types in our dataset, \textit{\lstinline{push}} (ranked 7th), \textit{\lstinline{workflow_dispatch}} to manually trigger a workflow (ranked 13), and \textit{\lstinline{pull_request}} (ranked 14)
are the most common.

To complement the frequency analysis, we also report the \emph{Median number of Occurrences per Workflow} (MOW) in Table~\ref{top_20}. It shows not only whether a construct is widely adopted across workflows, but also how often it is used within a single workflow.
For example, the fourth and fifth top constructs in Table~\ref{top_20} have a median of 5 occurrences per workflow. The third top construct has a MOW of $3$.
This indicates that maintainers typically use multiple Actions on multiple \textit{steps} in one or more \textit{jobs} that requires passing one or more parameters and providing a name for multiple steps.
Most of these multiply-used constructs reside at the step level, and suggest that the complexity of workflows can emerge not only from the diversity of constructs used, but also from the repeated use of the same construct within a workflow.

\summary{Most workflows use a small subset of the 197 \gha constructs.
Roughly half of them use between 9 and 14 constructs, with a skewed  distribution: a few constructs appear in nearly every workflow, whereas many constructs are rarely ever used. The most common constructs relate to using Action components and defining triggers.
Workflows also differ in how they use constructs. Some workflows rely on a wide variety of constructs but reuse them less frequently. Others stick to a smaller set of constructs but use them repeatedly.
}

\begin{table}[!t]
   
    \centering
    \begin{threeparttable}
 \footnotesize
        \caption{Top 15 %
        most frequent constructs based on their occurrences.}
    \label{top_20}
\begin{tabularx}{\columnwidth}{r X r r r}
          & \textbf{Construct} & \textbf{\#} & \textbf{MOW} & \textbf{\% WF} \\
          \midrule
            1 & \footnotesize\textit{\lstinline{name}} & 169K & --- & 99.1\% \\
            2 & \footnotesize\textit{\lstinline{jobs.<id>.runs-on}} & 281K & 1 & 95.4\% \\
            3 & \footnotesize\textit{\lstinline{jobs.<id>.steps[*].uses}} & 789K & 3 & 94.2\% \\
            4 & \footnotesize\textit{\lstinline{jobs.<id>.steps[*].with.<param>}} & 1,195K & 5 & 86.0\% \\
            5 & \footnotesize\textit{\lstinline{jobs.<id>.steps[*].name}} & 1,168K & 5 & 85.6\% \\
            6 & \footnotesize\textit{\lstinline{jobs.<id>.steps[*].run}} & 716K & 3 & 77.2\% \\
            7 & \footnotesize\textit{\lstinline{on.push.branches[*]}} & 93K & 1 & 42.2\% \\
            8 & \footnotesize\textit{\lstinline{jobs.<id>.name}} & 126K & 1 & 40.1\% \\
            9 & \footnotesize\textit{\lstinline{jobs.<id>.steps[*].env.<var>}} & 227K & 2 & 31.3\% \\
            10 & \footnotesize\textit{\lstinline{jobs.<id>.steps[*].if}} & 177K & 2 & 27.4\% \\
            11 & \footnotesize\textit{\lstinline{jobs.<id>.strategy.matrix.<var>}} & 97K & 1 & 27.0\% \\
            12 & \footnotesize\textit{\lstinline{jobs.<id>.steps[*].id}} & 103K & 1 & 26.8\% \\
            13 & \footnotesize\textit{\lstinline{on.workflow_dispatch}} & 42K & --- & 24.5\% \\
            14 & \footnotesize\textit{\lstinline{on.pull_request.branches[*]}} & 48K & 1 & 22.4\% \\
            15 & \footnotesize\textit{\lstinline{jobs.<id>.if}} & 55K & 1 & 19.0\% \\
        \bottomrule
        \end{tabularx}
          \begin{tablenotes}

         \item {--- for MOW signals constructs that can only be used once in a workflow.}
        \end{tablenotes}

      \end{threeparttable}
\vspace{-0.5cm}

\end{table}

\section{{$RQ_2$}: Which features are used in practice?}\label{sec:rq1b}

\begin{table*}[!t]
  \centering
  \footnotesize
  \begin{threeparttable}
    \caption{Usage statistics of GHA features in practice. Orange lines in the boxplots represent the median, black line the average.}
    \label{tab:features_usage}

\begin{tabularx}{0.9\textwidth}{
    r
    >{\raggedright\arraybackslash}p{2.6cm}
    S[table-format=2.1]
    X
    X
}

& \bf Feature & \multicolumn{1}{c}{\bf Usage Rate} &

\multicolumn{1}{c}{\bf Construct Coverage } &
\multicolumn{1}{c}{\bf Path-to-Construct Ratio} \\

\midrule
& & &
\raisebox{-0.5\height}{\includegraphics[width=\linewidth]{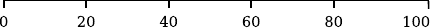}} & \raisebox{-0.5\height}{\includegraphics[width=\linewidth]{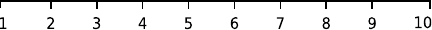}} \\

1 & naming & 99.9\% & \raisebox{-0.4\height}{\includegraphics[width=\linewidth]{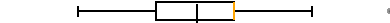}} & \raisebox{-0.4\height}{\includegraphics[width=\linewidth]{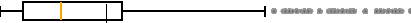}} \\

2 & triggers & 96.3\% & \raisebox{-0.3\height}{\includegraphics[width=\linewidth]{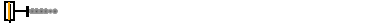}} & \raisebox{-0.3\height}{\includegraphics[width=\linewidth]{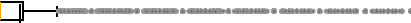}} \\

3 & context & 95.6\% & \raisebox{-0.3\height}{\includegraphics[width=\linewidth]{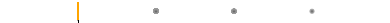}} & \raisebox{-0.4\height}{\includegraphics[width=\linewidth]{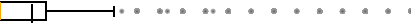}} \\
4 & Action reuse & 94.2\% & \raisebox{-0.3\height}{\includegraphics[width=\linewidth]{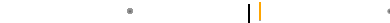}} & \raisebox{-0.3\height}{\includegraphics[width=\linewidth]{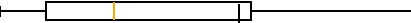}} \\
5 & commands & 77.5\% & \raisebox{-0.3\height}{\includegraphics[width=\linewidth]{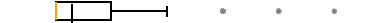}} & \raisebox{-0.3\height}{\includegraphics[width=\linewidth]{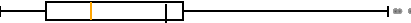}} \\
6 & environment variables & 45.6\% & \raisebox{-0.3\height}{\includegraphics[width=\linewidth]{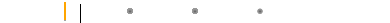}} & \raisebox{-0.3\height}{\includegraphics[width=\linewidth]{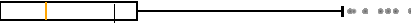}} \\
7 & job orchestration & 38.9\% & \raisebox{-0.3\height}{\includegraphics[width=\linewidth]{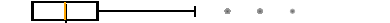}} & \raisebox{-0.3\height}{\includegraphics[width=\linewidth]{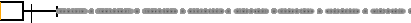}} \\
8 & permissions & 32.0\% & \raisebox{-0.3\height}{\includegraphics[width=\linewidth]{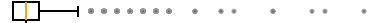}} & \raisebox{-0.3\height}{\includegraphics[width=\linewidth]{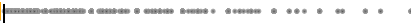}} \\
9 & matrix strategy & 30.7\% & \raisebox{-0.3\height}{\includegraphics[width=\linewidth]{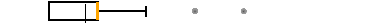}} & \raisebox{-0.3\height}{\includegraphics[width=\linewidth]{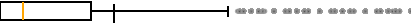}} \\
10 & step orchestration & 28.9\% & \raisebox{-0.3\height}{\includegraphics[width=\linewidth]{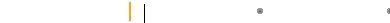}} & \raisebox{-0.3\height}{\includegraphics[width=\linewidth]{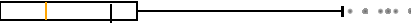}} \\
11 & workflow reuse & 12.4\% & \raisebox{-0.3\height}{\includegraphics[width=\linewidth]{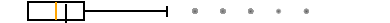}} & \raisebox{-0.3\height}{\includegraphics[width=\linewidth]{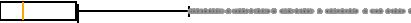}} \\

12 & deployment & 3.8\% & \raisebox{-0.3\height}{\includegraphics[width=\linewidth]{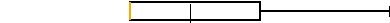}} & \raisebox{-0.3\height}{\includegraphics[width=\linewidth]{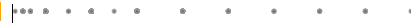}} \\

13 & container & 3.1\% & \raisebox{-0.3\height}{\includegraphics[width=\linewidth]{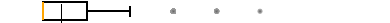}} & \raisebox{-0.3\height}{\includegraphics[width=\linewidth]{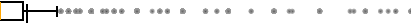}} \\
14 & services & 1.6\% & \raisebox{-0.3\height}{\includegraphics[width=\linewidth]{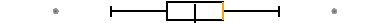}} & \raisebox{-0.3\height}{\includegraphics[width=\linewidth]{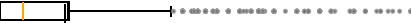}} \\ %

\bottomrule
\end{tabularx}
\end{threeparttable}
\vspace{-0.4cm}

\end{table*}

Based on the grouping of constructs into features (see $RQ_0$), we analyse how frequently workflows use features, and whether all constructs for a given feature are being used or only a subset of them.
\tab{tab:features_usage} summarises three characteristics for each feature across all 171,194 workflow snapshots:
the \emph{Usage Rate} (\ie the proportion of workflows using a feature), the distribution of its \emph{Construct Coverage},
and the distribution of the \emph{Path-to-Construct Ratio}.

Given a feature $f$ and a workflow $w$, the
 \textit{Construct Coverage} is computed by counting how many constructs from $f$ are used in $w$  divided by the total number of constructs associated to $f$.
The \textit{Path-to-Construct Ratio} is computed by dividing the number of paths in $w$ by the number of constructs associated to $f$ used in $w$.
\tab{tab:features_usage} caps this ratio at 10 ($95.6^{th}$ percentile) for readability.
As an example of how to interpret these distributions, consider the \emph{permissions} feature, encompassing 30 constructs (cf. Table~\ref{tab:features}) and used in 32\% of all workflows.
The \textit{Construct Coverage} boxplot reveals that few of the constructs are used in practice: for instance, half of the workflows (median value, orange line) use no more than 6.7\% (\ie 2 out of 30) \emph{permissions} constructs.
The \textit{Path-to-Construct Ratio} boxplot reveals a median of 1, implying that half of the workflows use \emph{permissions} constructs only once (\ie they define permissions once, at the level of the workflow, job or step, but do not redefine permissions multiple times).

\noindent\textbf{\emph{Usage Rate.}}
\tab{tab:features_usage} reveals that some features are more widely adopted than others.
Four features are used by more than 94\% of workflows.
99.9\% of all workflows use \emph{naming}, implying that assigning a name to a workflow, job, or step is a common practice for workflow maintainers.
The high usage rate of \emph{triggers} and \emph{context} is due to the necessity to define a trigger and runner for being able to execute a workflow. Finally, the high usage rate of \emph{Action reuse} (94.2\%) reflects that reusable Actions are crucial components of most workflows. The alternative way to define steps through custom scripts has a lower usage rate (feature \textsf{commands}, 77.5\%), suggesting that the wide availability of reusable Actions on the GitHub Marketplace %
makes \emph{Action reuse} a preferred practice.

In comparison, the \emph{workflow reuse} feature (12.4\%) is approximately eight times less used, suggesting that reusable workflows are either not well-known or too restrictive to be widely useful.
Five features are only moderately used: \emph{environment variables} (45.6\%), \emph{job orchestration} (38.9\%), \emph{permissions} (32.0\%), \emph{matrix strategy} (30.7\%), and
\emph{step orchestration} (28.9\%). They reflect more advanced workflow mechanisms that are generally useful to optimise and secure workflows, but less likely to be used by novice workflow maintainers.
Finally, three features are rarely used, namely \emph{deployment} (3.8\%), \emph{container} (3.1\%), and \emph{services} (1.6\%). They are considerably more specialised, explaining their low adoption rate. For example, the \emph{services} feature is used for hosting service containers for workflow jobs, which is only needed in specific situations.%

\noindent\textbf{\emph{Construct Coverage.}}
To analyse how features are used in practice, we examine the distributions of \emph{Construct Coverage} in \tab{tab:features_usage}. The table reveals that not all features are used in the same way.
For instance, the boxplots for \emph{naming}, \emph{Action reuse}, and \emph{services} show a high median construct coverage (60.0\%, 66.7\%, and 57.1\% respectively), implying that workflow maintainers using these features tend to use more constructs associated to them.
Since features \emph{naming} and \emph{Action reuse} contain a low number of constructs (5 and 3 respectively), it is easier to use more of their constructs and reach a high construct coverage.
The high median construct coverage for the rarely used \emph{services} feature (composed of 7 constructs) suggests that using it in workflows requires more complex configurations. %

It is worth noting that the construct coverage distributions of many features are skewed, indicating that while most workflows use only a few constructs, a small number of workflows tend to use a much larger number of constructs.
This suggests that some workflows leverage a much larger share of the available constructs, reflecting more advanced configurations.

\noindent\textbf{\emph{Path-to-Construct Ratio.}}
This metric captures how many times the constructs of a feature are used repeatedly within a workflow.
A ratio of 1 indicates that the constructs of the feature are used only once in the workflow, while a ratio above 1 indicates that the workflow uses some constructs multiple times.
\tab{tab:features_usage} reports on the distribution of this metric for all features. We obtain a median ratio of 1 for six features: \emph{context}, \emph{triggers}, \emph{permissions}, \emph{container}, \emph{job orchestration}, and \emph{deployment}.
Combined with the results of the previous feature-level metrics analysed, we can conclude that these features tend to be used in a simple way, using only 1 or 2 constructs without repetition.

In contrast, features like \emph{naming}, \emph{matrix strategy}, \emph{step orchestration}, \emph{environment variables}, \emph{workflow reuse}, and \emph{services} have a median value between 1 and 3, and features like \emph{Action reuse} and \emph{commands} have a median value above 3, which shows that some constructs associated to these features are used multiple times within a workflow.
Combined with the high median construct coverage for \emph{naming}, \emph{Action reuse}, and \emph{services}, this suggests that these features are used to their full extent (using most of their constructs), and that some of their constructs are used repeatedly in a workflow.

\summary{%
The 14 \gha \emph{features} show varying levels of adoption and usage patterns.
Features such as \emph{naming}, \emph{triggers}, \emph{context}, and \emph{Action reuse},
are widely adopted and often used repeatedly within a workflow.
Features such as \emph{commands}, \emph{environment variables}, and \emph{job orchestration} are moderately adopted and their constructs may be used multiple times within a workflow.
Specialized features such as \emph{workflow reuse}, \emph{deployment}, \emph{container} and \emph{services} are rarely adopted, with the latter two features showing high construct coverage, indicating that they are more complex and may require more advanced configurations.
}

\section{{$RQ_3$}: How does the usage of the GHA language evolve over time?}\label{sec:rq2}

While previous sections %
analysed construct and feature usage on recent workflow snapshots only,
this RQ aims to provide insights into how \gha usage has evolved over time.
For each month between July 2019 and August 2025, we materialized a snapshot of all workflows that were alive at that time (\ie created before that date and not yet deleted), resulting in 74 monthly snapshots for 243,008 unique workflows. We excluded 16,653 workflows with a lifespan of less than one month to ensure that we are analyzing workflows that had a chance to evolve over time.

\begin{figure*}[!t]
    \centering
    \includegraphics[width=0.8\linewidth]{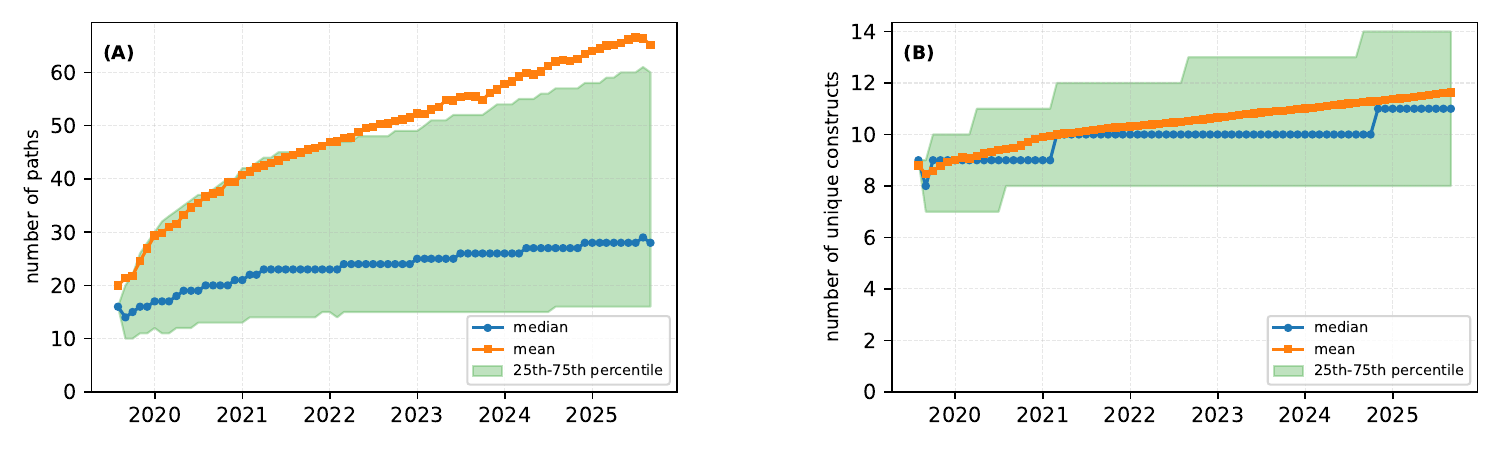}
    \caption{Monthly evolution of \textbf{(A)} \emph{\#Paths}, \textbf{(B)} \emph{\#Constructs}.}
    \label{fig:rq5_evolution_combined}
    \vspace{-0.2cm}

\end{figure*}

We analysed the evolution of workflow size using two different metrics: \emph{\#Paths} and \emph{\#Constructs}, aggregated by the mean, median, $25^{th}$ and $75^{th}$ percentile. Figures~\ref{fig:rq5_evolution_combined}-A and~\ref{fig:rq5_evolution_combined}-B show their monthly evolution. %
The two figures reveal a clear growth in the \emph{\#Paths} and \emph{\#Constructs}, indicating a trend towards increasing workflow size.
For instance, the median \emph{\#Paths} almost doubled over the observation period, from 16 to 31 paths, and the average \emph{\#Paths} has even tripled from 20 to 65 paths.
In contrast, the median \emph{\#Constructs} increased from 9 constructs in August 2019 to 11 constructs in August 2025, suggesting that workflows grow mostly in their number of paths, and much less in the constructs they use.

To confirm these trends, we applied the non-parametric Mann-Kendall test \cite{HenryMann1945NonparametricTests} to check for the presence of a monotonic trend in the \emph{\#Paths} and \emph{\#Constructs} metrics over time. The test results indicate a statistically significant increasing trend for both metrics ($p<0.001$), confirming that the observed growth in \emph{\#Paths} ($\tau=0.95$) and \emph{\#Constructs} ($\tau=0.76$) represents a consistent upward trend over the observation period.

\begin{figure}[!t]
    \centering
    \includegraphics[width=\linewidth]{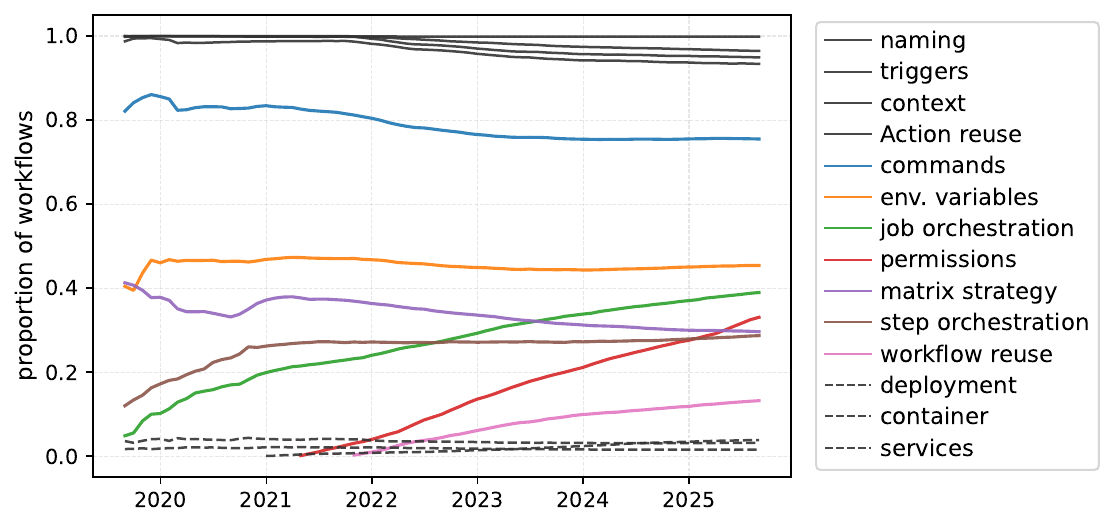}
    \caption{Monthly evolution of \emph{Feature Usage} in \gha workflows.}
    \label{fig:rq5_feature_usage_evolution}
    \vspace{-0.5cm}

\end{figure}

We also analysed the evolution of the \emph{\#Features} used in workflows. We found that this metric tends to remain relatively stable over time, with a median of 6 features used in a workflow during the entire observation period (mean ranged from 5.8 to 6.6). This suggests that while workflows are growing in size, they are not necessarily using more features of the \gha language.
To get a better understanding of the evolution of feature usage, we analysed the monthly proportion of workflows using each of the 14 features. The results are shown in Figure~\ref{fig:rq5_feature_usage_evolution}.
It reveals that the six most frequently used features have remained largely stable over time and did not change their relative order.
Likewise, the three least frequently used features have remained stable over time at a very low median usage rate (never exceeding 4\%).
Interestingly, the usage rate is consistently increasing over time for three features: \textit{job orchestration}, \textit{permissions}, and \textit{workflow reuse}. The last two were not part of the \gha language when it was first introduced in 2019 but were made available in 2021, illustrating the evolution of \gha language usage by the introduction and take-up of new language features, following \github's efforts for security hardening and DRY practices.
One can also observe that since the first appearance in our dataset of \textit{workflow reuse} (October 2021), the usage of some features such as \textit{context}, \textit{Action reuse} or \textit{commands} has begun to decrease slightly. This suggests a substitution effect where workflow maintainers are more and more resorting to reusable workflows to replace the functionalities previously implemented using existing features.

\summary{Workflow size has increased during the observation period by a factor of two to three in terms of number of paths, but only slightly in terms of number of constructs.
There is no observable growth in the number of features used, but specific features are seeing an increase in usage rate, in line with the introduction and promotion of new features in the \gha language.}

\section{{$RQ_4$}: To what extent does workflow size impact reliability and maintainability?}\label{sec:rq3}

Previous RQs focused on quantifying and analysing the evolution of workflow size in terms of \emph{\#Paths}, \emph{\#Constructs}, \emph{\#Features}, and \emph{Path-to-Construct-Ratio}.
Empirical analyses of traditional source code artefacts (\eg software programs) have used and studied a wide range of
code size metrics, such as the number of lines of code, the number of tokens, and the number of syntactic elements (e.g., functions, classes, packages) \cite{Fenton2014SoftwareMetrics}. Such size metrics have been established as strong, valid predictors of external quality features, specifically bug-proneness and maintenance effort \cite{Gil2017sizeValidity, Chowdhury2022maintainableMethodSize}.

Drawing on this analogy, we hypothesise workflow size to be related to reliability and maintainability issues. Our intuition is that larger workflows are more difficult to understand, test, and maintain, potentially leading to more frequent execution failures and higher maintenance effort.
To verify this hypothesis, we selected several metrics, used to measure \emph{maintainability} and \emph{reliability} of traditional software artifacts, and applied them to \gha workflows:

\noindent \textbf{\fr} is the percentage of workflow executions that fail. This metric is commonly used to measure the reliability of software artifacts \cite{IEEE982}.

\noindent \textbf{\commits} is the number of commits made to a workflow during a specific time window. This metric has already been used to measure the maintenance effort of \gha workflows \cite{Ghaleb2025FromFirstUse, Valenzuela2022}.
The use of a fixed, common time interval avoids biasing towards workflows that existed for a longer time, hence having more opportunity to be changed through commits.
It also avoids possible bias due to major changes that may have occurred over time in the language, technology or practices.

\noindent \textbf{\emph{Median Time to Repair (MTTR)}} measures
the median time required to restore a workflow to a working state after a failure. This is a standard measure of maintainability of software artifacts \cite{IEEE982}. A long time to repair can indicate that the workflow is more difficult to maintain.
To avoid bias towards repeated failures within the same workflow, we compute \mttr starting from the first observed failure of each workflow during a specific time window. We also exclude workflows that never recovered from the failure during this period.

\noindent \textbf{\aval} (a.k.a. uptime) is the proportion of time that a workflow is not
in a failed state \cite{IEEE982}.

\smallskip
To compute these metrics, we retrieved the execution results of the commits made to the workflows \changed{from the dataset of Cardoen \etal \cite{Cardoen2024}} over a six-month period from 20 January to 18 July 2025.
For each commit made to a workflow during this period, we used the GitHub REST API to collect the results of the run that were triggered by a change to the workflow file (either through a push or a pull request). By only considering commits that modify the workflow file contents and by focusing on the runs that were triggered by these changes, we reduce the effect of external factors that could affect the outcome of workflow execution (e.g., changes in the code base or external environment).
This resulted in 74,438 workflow run results associated with 13,915 workflows.
To understand the relationship between workflow size and reliability/maintainability, we used each of the four workflow size metrics to split all these workflows into three balanced groups reflecting \textbf{small}, \textbf{medium}, and \textbf{large} workflows. For example, based on the \emph{\#Paths} metric, the \textbf{small} group contained 4,730 workflows with ${\leq}29$ paths, the \textbf{medium} group contained 4,470 workflows with between 29 and 61 paths, and the \textbf{large} group contained 4,715 workflows with ${>}61$ paths. The discrete nature of the metric resulted in a slight variation in the number of workflows contained in each group.

\begin{table}[!t]
\begin{threeparttable}
\caption{Mann-Whitney U comparison of reliability and maintainability metrics between small (first third) and large (last third) workflows.}
  \label{tab:mwu_complexity}
\centering
\footnotesize
\setlength{\tabcolsep}{3pt}

\begin{tabular}{@{}l rr cccc@{}}

& \multicolumn{2}{c}{{workflow size}}
& \multicolumn{4}{c}{{Cliff's $\delta$ (\textbf{\emph{medium}}, \textbf{small} or negligible)}} \\
\cmidrule(lr){2-3} \cmidrule(lr){4-7}
{\bf Size metric}
  & {small $\leq$}
  & {large $>$}
  & \textbf{\emph{Failure R.}}
  & \textbf{\commits}
  & \textbf{\emph{Avail.}}
  & \textbf{\mttr} \\
\midrule
\#Paths                  & 29 & 61   & \textbf{0.228} & \textbf{\emph{0.374}} & \textbf{$-$0.247} & \textbf{0.174} \\
\#Constructs             & 12 & 16   & \textbf{0.151} & \textbf{\emph{0.411}} & \textbf{$-$0.175} & \textbf{0.186} \\
\#Features               & 7  & 8    & \textbf{0.172} & \textbf{\emph{0.332}} & \textbf{$-$0.190} & 0.135 \\
Path-to-Constr. & 2.31 & 4.02 & \textbf{0.236} & \textbf{0.300} & \textbf{$-$0.250} & 0.107  \\
\bottomrule
\end{tabular}

\end{threeparttable}
\vspace{-0.5cm}
\end{table}

To assess the effect of each size metric on the four reliability and maintainability metrics, we performed 16 non-parametric Mann-Whitney U tests to compare between the \textbf{small} and \textbf{large} groups.
For \mttr, we only considered 2,691 workflows as it is computed only on workflows that successfully recovered from an initial failure.
We applied a Benjamini-Hochberg adjustment %
 to control the false discovery rate due to multiple hypotheses testing \cite{Benjamini1995}, and set a significance level threshold of $\alpha{=}0.01$.
We could reject all null hypotheses with statistical significance: %
\textbf{large} workflows tend to have a higher \fr, more \commits, lower \aval, and a higher \mttr than \textbf{small} workflows.

\tab{tab:mwu_complexity} reports on the effect size of the differences, computed using Cliff's $\delta$ and interpreted based on Romano~\etal~\cite{romano2006exploring} (\ie  \emph{negligible} effect if $|\delta| < 0.147$, \emph{small} if $0.147 < |\delta| < 0.33$ and \emph{medium} effect if $0.33 \leq |\delta| < 0.474$).
One can observe \emph{{medium}} effect for 3 comparisons, \emph{small} effect for 11, and \emph{negligible} effect for only 2 \mttr comparisons.
These results suggest that larger workflows are more likely to fail, require more maintenance effort, and have lower availability.

\smallskip
While the Mann-Whitney U tests establish significant differences between \textbf{small} and \textbf{large} workflow groups, to gain deeper knowledge on the link between the increase of size metrics and workflow reliability and maintainability, we rely on Generalized Linear Modeling (GLM)
\cite{Nelder1972GeneralizedLinearModels}. It extends linear regression models via a link function to support response variables with a non-normal distribution. %
We excluded workflows with $<3$ %
runs, to avoid bias by workflows with very few runs.
Since \fr is %
bounded between 0 and 1, we use a {binomial logistic regression}, which is well-suited for this type of data.
Its effect size is reported using \emph{Odds Ratio} (OR).
For the discrete \commits metric, we use a {negative binomial regression}, which is suitable for overdispersed count data.
Its effect size is reported using \emph{Incidence Rate Ratio} (IRR).
\aval and \mttr could not be included in the GLM analysis, since their data distribution characteristics  (e.g., heavy right-skewness for \mttr and near-constant values for \aval) do not make them amenable to GLM.

Table~\ref{tab:GLM_general_size_metrics} reports the effect sizes of the regression analyses (OR for \fr, IRR for \commits), along with their 95\% confidence intervals. %
All regression analyses were found to be statistically significant ($\alpha{=}0.01$) after Benjamini-Hochberg adjustment~\cite{Benjamini1995}.
Effect sizes (OR and IRR) were all ${>}1$, indicating that an increase in the predictor variable is associated with an increase in the outcome variable.
Predictor variable \emph{\#Features} had the largest effect on both outcome variables. Adding one feature to a workflow is associated with an 18.9\% increase in the odds of failure and an 8.5\% increase in \commits.
The second best predictor variable was \emph{Path-to-Construct Ratio}, with an increase of one unit being associated with a 13\% increase in the odds of failure and a 7\% increase in \commits.
The predictive power of \emph{\#Constructs} and \emph{\#Paths} was considerably smaller.

This suggests that workflow maintainers should be cautious when creating or modifying large workflows, as they are more likely to face reliability and maintainability issues. This aligns with evidence from programming language research that suggests, for example, to keep method sizes small, since large methods are associated with more bug-proneness and lower maintainability \cite{Chowdhury2022maintainableMethodSize}.
In a \cicd environment where workflows execute frequently and are notoriously difficult to test and debug locally, even a 5\% cumulative increase in failure odds may result in significant wasted computing resources and manual troubleshooting efforts.

\summary{There is a statistically significant relation between workflow size and reliability and maintainability, with the number of unique features being the strongest predictor. Larger workflows are more likely to have more failures, more commits, longer downtime (\ie lower availability), and take more time to repair (\ie higher \mttr).
}

During our personal experience with maintaining workflows, we observed some features to be more error-prone to use and maintain than others.
For example, custom \emph{commands} in steps often require quite some technical expertise that makes them difficult to test and debug, whereas the alternative of relying on pre-built reusable Actions that have been tested by the community tends to reduce the likelihood of errors and failures.
To verify and generalise this hypothesis, we repeated the GLM regression analysis at the level of individual features, in order to determine which of these features are more likely to be associated with reliability and maintainability issues.
To ensure reliable estimations of effect size, we only focus on features that are used by more than 5\% and less than 95\% of the 13,915 considered workflows. Therefore, the universal features  \emph{naming} and \emph{triggers} (used by more than 95\%), and the highly specialized features \emph{services} and \emph{deployment} (used by fewer than 5\%) are excluded from this analysis.

\begin{table}[t]

\caption{Effect sizes and confidence intervals of GLM regression of workflow size metrics on \fr and \commits.
}

\label{tab:GLM_general_size_metrics}
\footnotesize
\centering
\begin{tabular}{lcccc}

& \multicolumn{2}{c}{\textbf{\fr}}
& \multicolumn{2}{c}{\textbf{\commits}} \\
\cmidrule(lr){2-3} \cmidrule(lr){4-5}
\textbf{Size metric}
& \textbf{OR} & \textbf{95\% CI}
& \textbf{IRR} & \textbf{95\% CI} \\
\midrule
  \#Paths         & 1.005 & 1.004--1.005 & 1.004 & 1.003--1.004 \\
  \#Constructs    & 1.024 & 1.018--1.029 & 1.048 & 1.042--1.054 \\
  \#Features      & 1.189 & 1.171--1.208 & 1.085 & 1.068--1.102 \\
  Path-to-Construct Ratio & 1.130 & 1.121--1.138 & 1.073 & 1.063--1.083 \\
 \bottomrule
\end{tabular}

\vspace{-.3cm}

\end{table}

Table \ref{tab:feature_results} reports two regression analyses to relate workflow features to both \fr and \commits of workflows.
The first analysis considers the presence (\ie the use) of a given feature in a workflow.
The second analysis considers a count variable indicating the number of paths added for a given feature in a workflow.
For example, workflows using shell \emph{commands} have over four times higher odds of failing compared to those that do not (OR=4.12) and are associated with a 15\% reduction in \commits (IRR = 0.85).
Adding additional command paths incrementally increases maintenance effort (IRR = 1.019 per path).
The presence of an \emph{environment variable} in a workflow increases the odds of failure by 84\% and increases the \commits by 35\%, suggesting that using environment variables can be error-prone and require more maintenance effort.
Adding one more environment variable increases the odds of failure and the \commits by 3\% each.
The presence of the \emph{step orchestration} feature in a workflow increases the odds of failure by 79\% and the \commits by 42\%, while adding one more path to this feature increases the odds of failure and the \commits by 2\% each.
Interestingly, the presence of \emph{Action reuse} in a workflow {decreases} the odds of failure by 28\% (OR=0.72) but {increases} maintenance effort by 58\% (IRR = 1.58).
This suggests that using \emph{Actions} instead of shell commands in workflows could be a more reliable option, although it comes with additional maintenance effort to keep these Actions up-to-date.

\summary{%
Relying on custom shell \emph{commands} significantly increases the odds of workflow failure,
whereas adopting \emph{reusable Actions} reduces the rate of failure but significantly increases ongoing maintenance effort (\text{\commits}) to manage external dependencies.}

\begin{table}[!t]
\centering
\begin{threeparttable}
\caption{Effect of feature presence and feature path count on \fr and \commits from GLM regression.}
\label{tab:feature_results}
\footnotesize
\setlength{\tabcolsep}{4pt}

\begin{tabular}{l cc cc}
& \multicolumn{2}{c}{\textbf{Feature Presence}} &
  \multicolumn{2}{c}{\textbf{\#Paths for the feature}} \\
\cmidrule(lr){2-3}\cmidrule(lr){4-5}

& \textbf{\fr} & \textbf{\commits} &
  \textbf{\fr} & \textbf{\commits} \\
\cmidrule(lr){2-2}\cmidrule(lr){3-3}\cmidrule(lr){4-4}\cmidrule(lr){5-5}

\textbf{Feature} & \textbf{OR} & \textbf{IRR} & \textbf{OR} & \textbf{IRR} \\
\midrule

Commands              & 4.12 & 0.85 & 1.018 & 1.019 \\
Environment variables & 1.84 & 1.35 & 1.032 & 1.031 \\
Step orchestration    & 1.79 & 1.42 & 1.023 & 1.025 \\
Job orchestration     & 1.66 & 1.20 & 1.023 & 1.047 \\
Matrix strategy       & 1.46 & 1.16 & 1.016 & 1.013 \\
Action reuse          & 0.72 & 1.58 & 1.012 & 1.013 \\
Permissions           & 0.70 & 1.42 & 0.838 & 1.086 \\
Context               & \multicolumn{1}{c}{---} & 1.58 & 1.099 & 1.064 \\
Container             & 1.58 & 1.32 & \multicolumn{1}{c}{---} & 0.391 \\
Workflow reuse        & 1.49 & 1.17 & \multicolumn{1}{c}{---} & \multicolumn{1}{c}{---} \\

\bottomrule
\end{tabular}

\begin{tablenotes}[flushleft]
\item All results except those marked as --- are statistically significant. The 95\% confidence intervals are available in the replication package.
\end{tablenotes}

\end{threeparttable}
\vspace{-0.3cm}
\end{table}

\section{Related work}\label{sec:related_work}
Many empirical studies have investigated the adoption, evolution, and challenges of CI/CD pipelines in open-source projects, primarily focusing on Travis CI.
These works highlighted the benefits of rapid feedback and pain points such as configuration bad smells, complex job matrices, and long build times \cite{Zampetti2021, Widder2019, gallaba2020tse, Zampetti2020, Elazhary_2022_BenefitsChallenges, hilton2016usage}.

Soon after its introduction, \gha became the dominant CI/CD service due to its deep integration with GitHub and a large marketplace of reusable Actions \cite{Golzadeh2022SANER}.
Studies have shown that \gha workflows are widely adopted but can become structurally complex \cite{decan2022gha}. They require continuous maintenance driven by bug fixes and CI/CD improvements \cite{Valenzuela2022}.
Expanding on this, empirical analyses have revealed that \gha \wfs are frequently modified, particularly in task specifications and configurations \cite{rostami2026GHAEvolution}. This continuous evolution provides evidence that \gha workflows follow Lehman’s laws of continuing growth and continuing change \cite{Lehman1999Metrics}.

Despite their proven benefits, maintaining reliable \gha workflows is challenging. Saroar and Nayebi~\cite{Saroar2023DevelopersPO} reported that YAML configurations are error-prone, while Zheng \etal~\cite{zheng2025ghafail} found that frequent workflow failures waste computing resources and require significant maintenance effort.
Looking at build outcomes specifically, Huang and Lin~\cite{Huang2026GHAReruns} investigated \gha workflow reruns caused by flakiness. They used structural metrics (e.g., lines of code and job count) alongside execution history as  features for machine learning models to predict execution outcomes. While their work focuses on predicting immediate rerun success, our study provides a fine-grained analysis of how specific usage of \gha language and size metrics impact workflow reliability and maintainability.

CI/CD configuration size directly impacts workflow maintainability.
Using a basic size metric (the number of directives) as a proxy for configuration complexity, Ghaleb \etal~\cite{ghaleb2025complexity} found \gha to be the second most complex CI/CD service.
In software engineering research, size and complexity metrics are well-established predictors of maintainability and defect-proneness.
Gil and Lalouche \cite{Gil2017sizeValidity} demonstrated that the validity of many code complexity metrics can be attributed to the confounding effect of size, while Chowdhury \etal \cite{Chowdhury2022maintainableMethodSize} provided empirical evidence linking size directly to future maintenance effort and bug-proneness.
Despite these insights, there is a lack of comprehensive understanding regarding how specific \gha language size and complexity metrics impact workflow reliability and maintenance effort.
Our study bridged this gap by investigating the relationship between multiple \gha workflow size metrics and workflow reliability and maintainability.%
\section{Threats to Validity}\label{sec:threats}

To discuss the threats to validity of our research we follow the structure recommended by Wohlin \etal~\cite{wohlin2012}.

\textbf{Construct validity} concerns the relation between theory and observed findings.
The dataset we relied on~\cite{Cardoen2024} includes only valid YAML workflow files that conform to the \gha JSON schema.
Still, some included workflows may have been disabled in practice.
We excluded workflows that use unsupported constructs (see \sect{sec:methodology}). Our construct catalog was derived from workflows alive as of August 2025, and may not cover constructs absent from all considered workflows. Given the scale of the dataset, however, the probability of missing a widely used construct is very low.  We are therefore confident that our results reflect representative usage patterns.\\
Finally, the execution results collected for $RQ_4$ pertain to commits that modified workflows. We cannot claim that failing workflow runs can solely be attributed to changes made to the workflow itself. External factors (e.g., dependency updates, third-party service outages) may also contribute.

\textbf{Internal validity} concerns factors that could influence observed outcomes independently of the intended measurements.
In \sect{sec:rq0} we manually mapped constructs to features. To mitigate bias, the authors independently proposed features and assigned constructs to them based on the official \gha documentation. Disagreements were then resolved by consensus.

\textbf{External validity} concerns the generalisability of our findings.
Our dataset is restricted to public \github repositories with at least 100 stars and 300 commits, criteria commonly used to exclude abandoned or experimental projects~\cite{Kalliamvakou2016}. Consequently, our findings may not generalise to smaller, less active repositories, nor to workflows hosted in private repositories.

%
%
\label{sec:conclusion}
Although the \gha language defines 197 constructs across 14 features, %
most workflows rely on only a small subset of them, typically between 9 and 14 constructs.
Features such as \emph{triggers}, \emph{naming}, \emph{context}, and \emph{Action reuse} dominate everyday practice, while more specialized features like \emph{containers}, \emph{services}, and \emph{workflow reuse} remain rare and tend to demand more complex configurations when adopted.
This imbalanced situation suggests that \gha usage comes with a steep learning curve, and that the cognitive burden on workflow maintainers scales with the breadth and rarity of the features they need to employ.
This raises questions about whether these features are niche by nature or whether their documentation and tooling support are insufficient.

Analysing the evolution of 243K workflows over a six-year period showed that workflow size doubled in terms of paths, though the number of constructs per workflow increased only slightly.
We also found that newly introduced or promoted features (such as \emph{permissions} or \emph{workflow reuse}) are gradually gaining adoption, reflecting the evolving nature of the language and its platform. This highlights the need for maintainers to continuously keep pace with language changes.

Critically, workflow size has measurable relationship with reliability and maintainability.
Our results indicate that larger workflows, especially those with more paths and a higher path-to-construct ratio, are associated with a higher failure rate, a lower availability, and a longer recovery time.
We also found that specific features tend to be associated with such reliability and maintainability risks. For instance, using \emph{commands} and \emph{environment variables} is associated with a higher failure rate, while using \emph{reusable Actions} is associated with a lower failure rate but a higher maintenance effort.
These findings suggest that certain language constructs may introduce more complexity or be more difficult to use correctly, leading to increased maintenance challenges and reliability issues.

Based on these insights, we outline actionable insights for different stakeholders.
Workflow maintainers should be aware that larger workflows, especially those using a wider variety of features and constructs, are more failure-prone and require more maintenance effort.
Understanding which features and constructs are more failure-prone can help practitioners make informed decisions about %
how to design their workflows to minimize reliability risks and maintenance effort.
While we cannot establish causality from our analysis, since there are many additional factors that could influence workflow reliability and maintainability, we still recommend practitioners to limit workflow size, potentially by modularizing their workflows through \emph{workflow reuse} mechanisms.

Tool builders can also benefit from our findings. For instance, linters could warn about overly long workflows and the use of specific features that are associated with higher failure rates or maintenance effort.
IDEs could provide suggestions to refactor large workflows, \eg to replace complex custom \emph{commands} with \emph{Actions} or to move and factorise long step sequences into reusable workflows.

Finally, researchers can use our results as a starting point for more in-depth quantitative and qualitative studies to address unanswered questions on the usage and complexity of \gha.
For instance, why are some features more failure-prone or require more maintenance effort than others? What are the ``atoms of confusion'' \cite{Gopstein2018AtomsOfConfusion} in workflows? Which best practices for using specific features and constructs could help reduce maintenance effort and reliability risks?
%
%

%

%
%
%
%

%
%

%
%
%

%
%

%
%
%
%

%
%
%
%

The replication package including all scripts, the construct to feature mapping, and instructions to reproduce the analysis
 is accessible on \url{https://figshare.com/s/224b0b49967f7ea8283a}.

\bibliographystyle{IEEEtran}
\bibliography{biblio}

@article{Onsori2026,
	author = {Onsori Delicheh, Hassan and Cardoen, Guillaume and Decan, Alexandre and Mens,Tom},
	doi = {10.48550/arXiv.2601.11299},
	journal = {arXiv preprint},
	title = {Automation and Reuse Practices in GitHub Actions Workflows: A Practitioner's Perspective},
	year = {2026}}

@article{Kalliamvakou2016,
	author = {Kalliamvakou, Eirini and Gousios, Georgios and Blincoe, Kelly and Singer, Leif and German, Daniel M. and Damian, Daniela},
	doi = {10.1007/s10664-015-9393-5},
	journal = {Empirical Software Engineering},
	number = {5},
	title = {An in-depth study of the promises and perils of mining {GitHub}},
	volume = {21},
	year = {2016}}

@article{Benjamini1995,
	author = {Yoav Benjamini and Yosef Hochberg},
	doi = {10.1111/j.2517-6161.1995.tb02031.x},
	issn = {00359246},
	journal = {Journal of the Royal Statistical Society. Series B (Methodological)},
	number = {1},
	pages = {289--300},
	publisher = {Oxford University Press},
	title = {Controlling the false discovery rate: A practical and powerful approach to multiple testing},
	volume = {57},
	year = {1995}}

@inproceedings{Cardoen2024,
	author = {Cardoen, Guillaume and Mens, Tom and Decan, Alexandre},
	booktitle = {Int'l Conf. Mining Software Repositories},
	doi = {10.1145/3643991.3644867},
	pages = {677--681},
	publisher = {ACM},
	title = {A dataset of {GitHub Actions} workflow histories},
	year = {2024}}

@article{ghaleb2025complexity,
	author = {Ghaleb, Taher and Abduljalil, Osamah and Hassan, Safwat},
	doi = {10.1145/3736758},
	journal = {ACM Trans. Softw. Eng. Methodol.},
	month = may,
	publisher = {ACM},
	title = {{CI/CD} Configuration Practices in Open-Source {Android} Apps: An Empirical Study},
	year = {2025}}

@article{zheng2025ghafail,
	author = {Zheng, Lianyu and Li, Shuang and Huang, Xi and Huang, Jiangnan and Lin, Bin and Chen, Jinfu and Xuan, Jifeng},
	doi = {10.1145/3749371},
	journal = {ACM Trans. Softw. Eng. Methodol.},
	month = jul,
	publisher = {ACM},
	title = {Why Do {GitHub Actions} Workflows Fail? {An} Empirical Study},
	year = {2025}}

@inproceedings{decan2022gha,
	author = {Decan, A. and Mens, T. and Rostami Mazrae, P. and Golzadeh, M.},
	booktitle = {Int'l Conf. Software Maintenance and Evolution},
	doi = {10.1109/ICSME55016.2022.00029},
	organization = {IEEE},
	title = {On the use of {GitHub Actions} in software development repositories},
	year = {2022}}

@inproceedings{Golzadeh2022SANER,
	author = {Golzadeh, Mehdi and Decan, Alexandre and Mens, Tom},
	booktitle = {Int'l Conf. Software Analysis, Evolution and Reengineering},
	doi = {10.1109/SANER53432.2022.00084},
	organization = {IEEE},
	title = {On the rise and fall of {CI} services in {GitHub}},
	year = {2022}}

@inproceedings{hilton2016usage,
	author = {Hilton, Michael and Tunnell, Timothy and Huang, Kai and Marinov, Darko and Dig, Danny},
	booktitle = {Int'l Conf. Automated Software Engineering ({ASE})},
	organization = {IEEE},
	pages = {426--437},
	title = {Usage, costs, and benefits of continuous integration in open-source projects},
	year = {2016}}

@inproceedings{romano2006exploring,
	author = {Romano, Jeanine and Kromrey, Jeffrey D and Coraggio, Jesse and Skowronek, Jeff and Devine, Linda},
	booktitle = {Annual Meeting of the Southern Association for Institutional Research},
	title = {Exploring methods for evaluating group differences on the {NSSE} and other surveys: Are the t-test and {Cohen's} d indices the most appropriate choices?},
	year = {2006}}

@inproceedings{Saroar2023DevelopersPO,
	author = {Saroar, S. G. and Nayebi, M.},
	booktitle = {Int'l Conf. Evaluation and Assessment in Software Engineering},
	doi = {10.1145/3593434.3593475},
	title = {Developers' perception of {GitHub Actions}: A survey analysis},
	year = {2023}}

@inproceedings{Valenzuela2022,
	author = {Valenzuela-Toledo, Pablo and Bergel, Alexandre},
	booktitle = {Int'l Conf. Software Analysis, Evolution and Reengineering ({SANER})},
	doi = {10.1109/saner53432.2022.00026},
	publisher = {IEEE},
	title = {Evolution of {GitHub Action} workflows},
	year = {2022}}

@inproceedings{Widder2019,
	author = {Widder, David Gray and Hilton, Michael and K\"{a}stner, Christian and Vasilescu, Bogdan},
	booktitle = {Joint Meeting on European Software Engineering Conference and Symposium on the Foundations of Software Engineering},
	doi = {10.1145/3338906.3338922},
	isbn = {9781450355728},
	publisher = {ACM},
	title = {A Conceptual Replication of Continuous Integration Pain Points in the Context of {Travis CI}},
	year = {2019}}

@book{wohlin2012,
	author = {Wohlin, C. and Runeson, P. and H\"ost, M. and Ohlsson, M. C. and Regnell, B. and Wessl\'en, A.},
	isbn = {3642290434},
	publisher = {Springer},
	title = {Experimentation in Software Engineering},
	year = {2012}}

@article{Zampetti2020,
	author = {Zampetti, Fiorella and Vassallo, Carmine and Panichella, Sebastiano and Canfora, Gerardo and Gall, Harald and Di Penta, Massimiliano},
	doi = {10.1007/s10664-019-09785-8},
	journal = {Empirical Software Engineering},
	title = {An empirical characterization of bad practices in continuous integration},
	volume = {25},
	year = {2020}}

@inproceedings{Zampetti2021,
	author = {Zampetti, Fiorella and Geremia, Salvatore and Bavota, Gabriele and Di Penta, Massimiliano},
	booktitle = {Int'l Conf. Software Maintenance and Evolution ({ICSME})},
	doi = {10.1109/ICSME52107.2021.00048},
	title = {{CI/CD} pipelines evolution and restructuring: A qualitative and quantitative study},
	year = {2021}}

@misc{githubActionsDocs,
	author = {GitHub},
	howpublished = {\url{https://docs.github.com/en/actions/using-workflows/workflow-syntax-for-github-actions}},
	note = {Accessed: 2025-09-22},
	title = {Workflow syntax for {GitHub Actions}},
	year = {2025}}

@book{gini1912Variabilit,
	adsurl = {https://ui.adsabs.harvard.edu/abs/1912vamu.book.....G},
	author = {{Gini}, C.},
	publisher = {Tipogr. di P. Cuppini},
	title = {{Variabilit{\`a} e mutabilit{\`a}}},
	year = 1912}

@article{mernik2005DSL,
	author = {Mernik, Marjan and Heering, Jan and Sloane, Anthony M},
	doi = {10.1145/1118890.1118892},
	journal = {ACM computing surveys},
	number = {4},
	publisher = {ACM},
	title = {When and how to develop domain-specific languages},
	volume = {37},
	year = {2005}}

@inproceedings{Lehman1999Metrics,
	author = {Lehman, M. M. and Ramil, J. F. and Wernick, P. D. and Perry, D.~E. and Turski, W.~M.},
	booktitle = {Proceedings of the IEEE Metrics Symposium (Metrics '99)},
	note = {FEAST/1 Technical Report (public PDF)},
	title = {Metrics and Laws of Software Evolution---The Nineties View},
	year = {1999}}

@article{Elazhary_2022_BenefitsChallenges,
	author = {Elazhary, Omar and Werner, Colin and Li, Ze Shi and Lowlind, Derek and Ernst, Neil A. and Storey, Margaret-Anne},
	doi = {10.1109/tse.2021.3064953},
	issn = {2326-3881},
	journal = {IEEE Transactions on Software Engineering},
	month = jul,
	number = {7},
	pages = {2570--2583},
	publisher = {IEEE},
	title = {Uncovering the Benefits and Challenges of Continuous Integration Practices},
	volume = {48},
	year = {2022}}

@article{gallaba2020tse,
	author = {Keheliya Gallaba and Shane McIntosh},
	journal = {IEEE Transactions on Software Engineering},
	number = {1},
	pages = {33--50},
	title = {{Use and Misuse of Continuous Integration Features: An Empirical Study of Projects that (mis)use Travis CI}},
	volume = {46},
	year = {2020}}

@inproceedings{Ghaleb2025FromFirstUse,
	author = {Chopra, Nitika and Ghaleb, Taher A.},
	booktitle = {Int'l Conf. Software Maintenance and Evolution ({ICSME})},
	doi = {10.1109/ICSME64153.2025.00078},
	pages = {773-778},
	publisher = {IEEE},
	title = {From First Use to Final Commit: Studying the Evolution of Multi-{CI} Service Adoption},
	year = {2025}}

@misc{IEEE982,
	address = {New York, NY, USA},
	author = {{IEEE}},
	institution = {{IEEE}},
	number = {982.2-1988},
	title = {{IEEE} Guide for the Use of {IEEE} Standard Dictionary of Measures to Produce Reliable Software},
	type = {Standard},
	year = {1988}}

@book{Fenton2014SoftwareMetrics,
	author = {Fenton, Norman and Bieman, James},
	doi = {10.1201/b17461},
	isbn = {9780429106224},
	month = {10},
	title = {Software Metrics: A Rigorous and Practical Approach, Third Edition},
	year = {2014}}

@article{Nelder1972GeneralizedLinearModels,
	author = {J. A. Nelder and R. W. M. Wedderburn},
	doi = {10.2307/2344614},
	journal = {Journal of the Royal Statistical Society. Series A (General)},
	number = {3},
	pages = {370--384},
	publisher = {Oxford University Press},
	title = {Generalized Linear Models},
	urldate = {2026-02-19},
	volume = {135},
	year = {1972}}

@article{Gil2017sizeValidity,
	author = {Gil, Yossi and Lalouche, Gal},
	doi = {10.1007/s10664-017-9513-5},
	journal = {Empirical Software Engineering},
	month = {10},
	pages = {1-27},
	title = {On the correlation between size and metric validity},
	volume = {22},
	year = {2017}}

@inproceedings{Chowdhury2022maintainableMethodSize,
	author = {Chowdhury, Shaiful Alam and Uddin, Gias and Holmes, Reid},
	booktitle = {Int'l Conf. Mining Software Repositories ({MSR})},
	doi = {10.1145/3524842.3527975},
	pages = {252--264},
	publisher = {ACM},
	title = {An empirical study on maintainable method size in {Java}},
	year = {2022}}

@article{rostami2026GHAEvolution,
	author = {Rostami Mazrae, Pooya and Decan, Alexandre and Mens, Tom and Wessel, Mairieli},
	doi = {10.1016/j.jss.2026.112824},
	issn = {0164-1212},
	journal = {Journal of Systems and Software},
	title = {An empirical study of the evolution of {GitHub Actions} workflows},
	volume = {236},
	year = {2026}}

@article{Huang2026GHAReruns,
	author = {Huang, Jiangnan and Lin, Bin},
	doi = {10.1145/3795771},
	issn = {1049-331X},
	journal = {ACM Trans. Softw. Eng. Methodol.},
	month = feb,
	publisher = {ACM},
	title = {On the Reruns of {GitHub Actions} Workflows},
	year = {2026}}

@article{HenryMann1945NonparametricTests,
 ISSN = {00129682, 14680262},
 URL = {http://www.jstor.org/stable/1907187},
 author = {Henry B. Mann},
 journal = {Econometrica},
 number = {3},
 pages = {245--259},
 publisher = {[Wiley, Econometric Society]},
 title = {Nonparametric Tests Against Trend},
 urldate = {2026-03-05},
 volume = {13},
 year = {1945}
}

@inproceedings{Gopstein2018AtomsOfConfusion,
author = {Gopstein, Dan and Zhou, Hongwei Henry and Frankl, Phyllis and Cappos, Justin},
title = {Prevalence of confusing code in software projects: atoms of confusion in the wild},
year = {2018},
isbn = {9781450357166},
publisher = {Association for Computing Machinery},
address = {New York, NY, USA},
url = {https://doi.org/10.1145/3196398.3196432},
doi = {10.1145/3196398.3196432},
booktitle = {Proceedings of the 15th International Conference on Mining Software Repositories},
pages = {281–291},
numpages = {11},
location = {Gothenburg, Sweden},
series = {MSR '18}
}

\end{document}